# Measuring and Modeling Physics Students' Conceptual Knowledge Structures Through Term Association Times

**Ian D. Beatty**

*University of Massachusetts, Amherst, MA, USA[1]*

**William J. Gerace**

*University of Massachusetts, Amherst, MA, USA*

**Robert J. Dufresne**

*University of Massachusetts, Amherst, MA, USA*

Traditional problem-based exams are not efficient instruments for assessing the "structure" of physics students' conceptual knowledge or for providing diagnostically detailed feedback to students and teachers. We present the *Free Term Entry* task, a candidate assessment instrument for exploring the connections between concepts in a student's understanding of a subject. In this task, a student is given a general topic area and asked to respond with as many terms from the topic area as possible in a given time; the "thinking time" between each term-entry event is recorded along with the response terms. The task was given to students from two different introductory physics classes. Response term thinking times were found to correlate with the strength of the association between two concepts. In addition, sets of thinking times from the task show distinct, characteristic patterns which might prove valuable for student assessment. We propose a quantitative dynamical model named the *Matrix Walk Model* which is able to match many aspects of the observed data. One particular feature of the data — a distinct "spike" superimposed on the otherwise log-normal distribution of most thinking time sets — has not been fit. The spike, other patterns observed in the data, and the proposed phenomenological model could all benefit from a grounding in cognitive theory.

---

[1] Address for correspondence: Dr. Ian Beatty, Physics Department, Lederle Graduate Research Tower 417A, University of Massachusetts, Amherst MA 01003-4525 USA. E-mail: beatty@physics.umass.edu. Home page: http://umperg.physics.umass.edu/

# I. Introduction

"Assessment drives pedagogy" is an oft-heard phrase in educational research circles. It derives from the tendency of grade-conscious students to optimize their learning for exam results, and of results-conscious teachers to optimize their curriculum for class performance on assessments. It holds true for the style as well as the content of assessments: if tests target rote knowledge, students will perceive that as the objective of instruction; if they target conceptual reasoning and transfer, students will more likely focus on those.

Therefore, assessment is presently a lively topic of educational research (Nichols, Chipman, & Brennan, 1995). Much of this research is devoted to developing *cognitively diagnostic*, *formative assessments*. "Cognitively diagnostic" means the assessments can be used to characterize the state of knowledge of individual students with enough detail to guide students' subsequent learning efforts and teachers' subsequent interventions, as opposed to merely characterizing the learning of a population of students (e.g., for curriculum evaluation) benchmarking students' gross level of mastery (e.g., for placement decisions). "Formative" means the assessments are used in an ongoing way during instruction to guide and enhance teaching and learning, rather than after instruction to evaluate success.

Cognitively diagnostic assessment requires two foundations: instruments that can probe and gather data on relevant features of a student's state of knowledge, and a model of topic knowledge and learning by which the data may be interpreted. Development of these two foundations is necessarily interdependent, because any proposed probe instrument can only be justified through the interpretability of the data it yields, and interpretation requires a model; and a model is justified by its ability to explain observed data. This chicken-and-egg relationship is common to all young research fields.

Acknowledging this, we have attempted to simultaneously develop assessment methods for probing introductory physics students' state of knowledge at a fundamental level, and a complementary quantitative model of conceptual knowledge. In a previous paper (Beatty & Gerace, 2001) we proposed a family of computer-based tasks as assessment instruments, and presented data suggesting that the tasks are sensitive to relevant aspects of students' *conceptual knowledge structure* (CKS). In this paper we focus on one of those tasks, analyzing in more detail the data it can provide. In addition, we suggest a simple quantitative dynamical model of CKS and how it is accessed which can "explain" the observed data.

It is not our intent here to suggest new research tools for basic cognitive science research, nor to propose a mature, ready-to-use assessment instrument, nor to present a complete or fundamental cognitive model. Rather, it is to explore the potential of a new (in this context) assessment approach for eliciting information about students' CKS, and to demonstrate how the data produced can be quantitatively modeled. That is, we are attempting a "proof-of-concept" venture. Our results are for the domain of introductory physics, but we see no reason why they might not be applicable to other educational levels and other subjects in which conceptual understanding is important.

Section II motivates and describes the *Free Term Entry* assessment task. Section III phenomenologically describes data obtained via the task. Section IV defines the *Matrix Walk Model* and evaluates its ability to reproduce experimentally observed data. Section V discusses the results.

## II. Probing Conceptual Knowledge Structure

It is useful to represent physics knowledge as divided into four general categories: *conceptual* knowledge, *operational and procedural* knowledge, *problem-state* knowledge, and *strategic* knowledge (Chi, Feltovich, & Glaser, 1981; Dufresne, Leonard, & Gerace, 1992; Gerace, 1992; Larkin, 1979; Leonard, Gerace, Dufresne, & Mestre, 1999). Conceptual knowledge includes concepts and general ideas such as "work" and "conservation of energy" and the associations between them. Operational and procedural knowledge includes equations, operations, and procedures such as defining a coordinate system or identifying all forms of energy present in a situation. Problem-state knowledge includes recognition of previously encountered problems and memory of relevant facts about their solution. Strategic knowledge links elements of the other categories into higher-level elements that guide the entire problem-solving process, like a stratagem for determining the principles to apply to a problem or methods of error-checking an answer.

Research results have revealed that for expert-like behaviors such as qualitative analysis, reasoning, and knowledge transfer to novel situations, *conceptual* knowledge is especially important (for a review, see Bransford, Brown, & Cocking, 1999). In particular, the organization and not just the content of an individual's conceptual knowledge is crucial: experts have *contextually appropriate access* to knowledge; it is the *structure of interconnections* between knowledge elements which allows such access; and experts' knowledge is structured around key *principles* (Chi et al., 1981; Hardiman, Dufresne, & Mestre, 1989; and work reviewed in Mestre & Touger, 1989; Redish, 1994; Zajchowski & Martin, 1993).

Therefore, we seek methods to assess as directly as possible students' conceptual knowledge structure (CKS) — the content of, organization of, and interconnections within conceptual knowledge. In addition to a sufficiently detailed model of conceptual knowledge and its access, this requires assessment instruments (probes) sensitive to elements of CKS. Conventional problem-based assessments are insufficient, because they indicate whether a student can do a problem but not why she is able or unable to. A student might fail to solve a particular problem for many different reasons: for example, failure to interpret the problem situation as intended, incorrect or insufficient physical intuition, ignorance of the necessary principle, misapplication of the correct principle, algebraic or numerical error, general cognitive overload and confusion, or failure to answer the precise question being asked. If a student's written solutions are hand-graded, it might be possible to infer the nature of the student's mistake; doing this for many students and for many problems per student is impractical. And even then, identifying the nature of the student's mistake does not necessarily indicate the particular failure of knowledge, conceptual or otherwise, causing it.

A general, qualitative model of conceptual knowledge structure which has seen some success is the *semantic map* (Quillian, 1968; Woods, 1975): concepts are represented as labeled nodes like "energy" and "movement", and the relationships between concepts as labeled, oriented links like "is an example of" and "causes". This perspective led to the development of *concept mapping* as a research and assessment tool, in which students are asked to construct or to complete a semantic map for some topic area (Novak & Gowin, 1984). The resulting diagram is interpreted as a description, perhaps partial, of the subject's conceptual knowledge structure. Many variants of the task have been investigated. In general, investigators find it to be useful for research, instruction, and small-scale assessment, but impractical for widespread use (Rice, Ryan, & Samson, 1998; Ruiz-Primo & Shavelson, 1996; Young, 1993).

In accordance with the semantic map nodes-and-links picture of conceptual knowledge, we have devised some computer-based tasks with which to obtain quantitative experimental data on the interconnections (links) between physics students' conceptual knowledge elements (Beatty, 2000; Beatty & Gerace, 2001). This paper will consider one of those tasks, the *Free Term Entry* (FTE) task.

FTE is a sort of "quasi-free association" in which subjects are given a topic area and asked to respond with a sequence of terms they consider associated with the topic area. A subject is presented with a general topic area such as "introductory mechanics" or "the material covered in the physics course you just completed". The subject is asked to think of terms that he associates with this topic area, and to type these terms into a computer, spontaneously and without forethought, as they come to mind. The response terms and associated timing information are recorded. The task is not truly "free" association since the subject is asked to eschew terms outside of the specified topic and to avoid entering any term more than once. The task lasts for a specified total duration, perhaps 15 to 45 minutes, and subjects typically enter between 40 and 120 terms.

Figure 1 shows the dialog box presented to the subject for term entry. When the subject finishes typing a term, he presses the keyboard's "return" key or clicks the "Enter" button and the typed term disappears, leaving the typing box ready for another term. The "Clear" button erases any currently-typed text without entering it, and the "Pause" button allows the subject to interrupt the task briefly (to ask a question, for example). Subjects are generally discouraged from using the pause facility. The data recorded consists of the list of response terms in the order they are entered, the time at which the typing of each term begins (defined as the time at which the first character is typed into an empty box), and the time at which the subject presses "return" to enter each term.

The FTE task has been designed to focus on *terms* rather than on equations, propositions, or other kinds of entities because a term seems to be the closest accessible approximation to a "conceptual building block". It has proven difficult to rigorously define *term*. When instructing subjects, a term is loosely defined to be one or perhaps two or three words describing one concept, idea, or thing. Some examples of terms drawn from introductory mechanics are "kinematics", "Newton's first law", "pulley" and "problem-solving". Statements like "energy is conserved in an elastic collision" are not considered to be terms, but rather propositions involving multiple terms and their relationship. "Conservation of energy", on the other hand, would be accepted as a term, since it serves as the name of a physics idea.

Figure 1: Term entry dialog box for the Free Term Entry (FTE) task.

The task is intended to explore the space of terms constituting a subject's "active vocabulary" of concepts for the topic area by allowing subjects to display their own choice of terms, rather than prompting them with a preselected set. The task is also intended to elicit subjects' spontaneous associations rather than considered, reflective decisions (e.g., for concept map drawing). It is hypothesized that the duration of pauses between term entries, and the grouping of term entries into clusters separated by longer fallow periods, can reveal information about what terms a subject associates closely. Since the list of terms and times comprising an FTE data set forms a one-dimensional series, and the space of conceptual knowledge elements and their interconnections requires at least two dimensions to represent (for example, as a matrix of connection strengths), the FTE task can never provide a complete probe of a subject's conceptual interconnections. Nevertheless, it might prove to be a useful sampling of those interconnections.

Ultimately, for practical assessment, we do not expect the FTE task to be used in isolation but rather in conjunction with complementary tasks described elsewhere (Beatty, 2000; Beatty & Gerace, 2001), and perhaps with as-yet-undeveloped tasks. In this paper it is considered alone, however, for the purpose of understanding the data it provides and constructing a model.

## III. Phenomenology

This section will describe the data obtained by the Free Term Entry task, focusing on aspects that will shape the modeling efforts of Section IV. Subsection A describes the data and indicates how it was acquired. Subsection B presents evidence that the timing information captured by the task does, as intended, reveal conceptual associations in the student's mind. Subsections C and D display phenomenological patterns observed in the data.

**A. The Studies and Data**

Two formal studies were conducted in which student volunteers from introductory physics courses at the University of Massachusetts Amherst were given the FTE task to complete. One study was conducted with 17 students drawn from General Physics II (for science and engineering majors) during Fall 1997; the other was conducted with 16 students from General Physics I (for science and engineering majors) during Spring 1999. In both cases, volunteers were solicited from the class by public announcement, with financial compensation offered. All subjects were proficient in spoken English.

For the Physics II study, subjects were selected after the third of three evening exams, chosen from the pool of volunteers such that the set of subjects' three-exam grade averages spanned the range from "D" through "A". The FTE task was given after the last day of classes and before the final exam. The specified domain area was "the subject material covered by the course all semester" (i.e., introductory thermodynamics, electricity & magnetism), and the duration of the task was approximately 50 minutes.

For the Physics I study, subjects were selected after the first evening exam, and chosen to get a reasonably uniform distribution of exam grades from "C" through "A". Subjects participated in a 15-minute session every week for nine weeks, followed by a 90-minute session during the last week of classes. The FTE task was given only once, during the final session; the specified domain area was "the material covered in [your physics class]" (introductory mechanics) and the task duration was 30 minutes. Other tasks not relevant to this paper comprised the rest of the study.

In addition, several "expert subjects" (physics faculty and graduate students) were used as preliminary subjects. After completing the task, they were asked to reflect upon their mental activity during it. Their introspections helped guide hypothesis formation and interpretation of the observations.

The raw data captured for each subject's performance on the FTE task was a list of the terms entered, in the order entered, along with the time at which the first letter of each term was typed (*start time*) and the time at which the return key was pressed to complete the term (*enter time*). A term's *typing time* is defined to be the difference between the term's enter and start times, indicating how long the subject spent typing the term. A term's *thinking time* is defined to be the difference between the term's start time and the previous term's enter time, indicating how much time passed between the two terms while the subject was not typing.

**B. Thinking Time Probability Distribution**

We define a term's *index* to be 1 if it was the first term entered in a subject's FTE response set, 2 if it was the second entered, etc. Figure 2 shows a plot of each entered term's thinking time vs. its index for a representative FTE data set, that of subject 01 in the Physics I study. The thinking times appear randomly distributed inside an envelope that increases with term index, with short times occurring throughout the task and longer times more likely to occur late in the task.

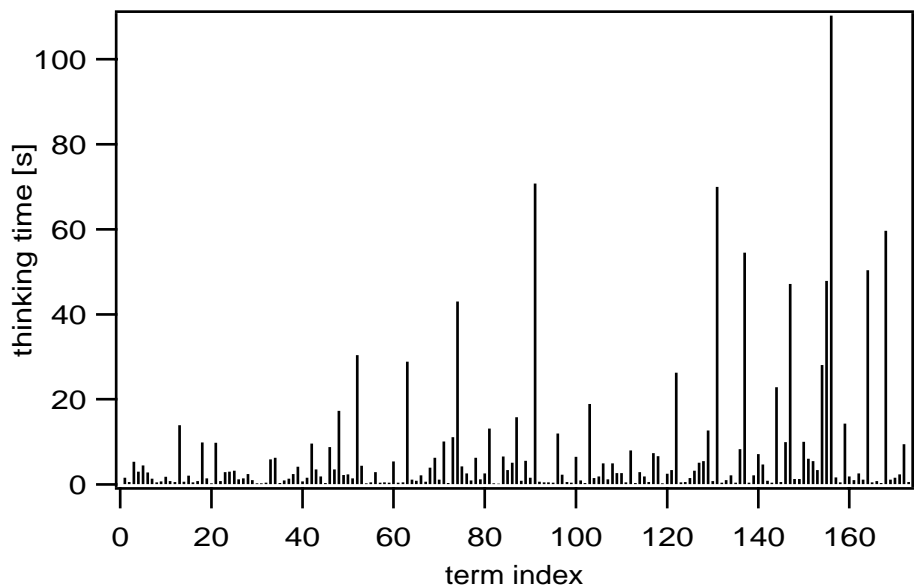

Figure 2: Thinking time vs. term index for subject 01's FTE data from the Physics I study.

Disregarding for now the systematic trend of increasing times with term index, each subject's set of thinking times can be analyzed as if it were a set of uncorrelated values drawn from a random distribution, and the nature of that distribution can be explored. Figure 3 shows a histogram of the natural logarithms of the thinking times for the data depicted in Figure 2. The natural logarithm of the times has been used because short times are far more common than long times, and a linear scale that included the longest times would lose detail for the short times. The distribution appears approximately normal (Gaussian), indicating that the distribution of thinking times themselves is approximately log-normal. (A random variable obeys a *log-normal* distribution if its logarithm obeys a normal distribution.) Deviating from normality, the histogram has a slight skew to the right and a significant spike superimposed on the leading edge.

These three features — normality, skew, and spike — are in fact common to almost all of the data sets collected, except where obscured by statistical noise in sets with relatively few response terms. Examination of the histograms for all 33 subjects in both studies shows that most of the histograms display a pronounced spike on the left edge of a broad peak, and very few of the histograms do not have at least a rudimentary bump there. The evidence for a "spike plus Gaussian" distribution of thinking time logarithms is strong, and this is a distinguishing feature of the data for candidate models to reproduce. The thinking times describing the spike and peak locations and the fraction of counts ascribable to the spike might eventually be interpretable as characteristic cognitive measures of the subject. In addition, with a sufficient model, best-fit parameters obtained from a student's data set by fitting a log-normal or other distribution to the thinking times might be informative.

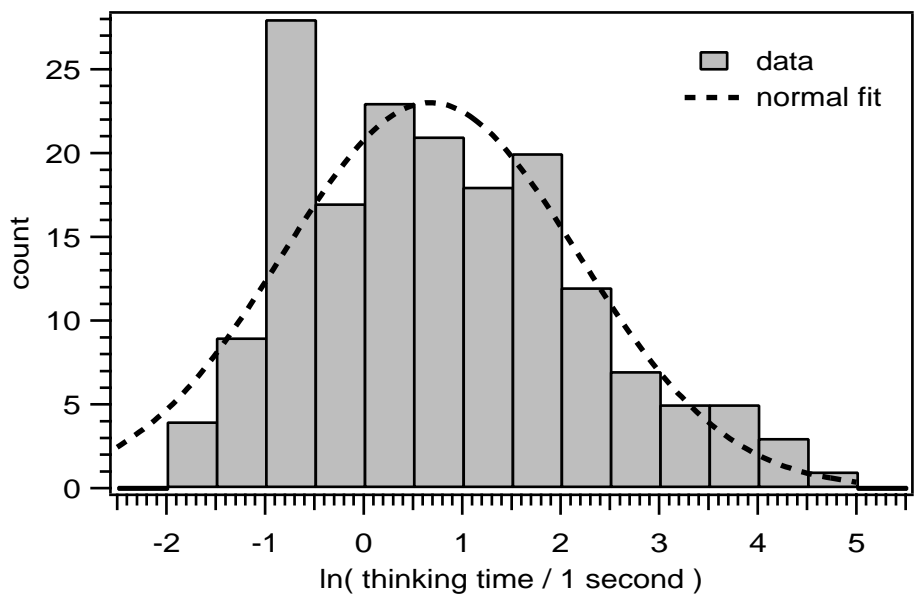

Figure 3: Histogram of the natural logarithms of the thinking times for subject 01's FTE data from the Physics I study, with best-fit normal probability distribution function (PDF).

**C. Temporal Correlations**

The previous subsection examined FTE thinking times as if they were uncorrelated numbers drawn from a random distribution. Such a description is incomplete: the times form a well-ordered sequence from the beginning to the end of the task, and both overall trends and correlations between neighboring values are likely. Examining Figure 2 and equivalent plots for other subjects, two global patterns are evident:

- As the task progresses, long thinking times become more frequent, and the long times that appear tend to have much larger values; and
- Short times continue to occur throughout the task.

Both patterns make intuitive sense: term entries become increasingly sparse later in the task, with longer pauses, because subjects have already entered most of their readily accessible terms and have to think hard to recall additional terms; and term entries tend to occur in clusters separated by short times because when a subject thinks of a term, it often suggests other connected terms. Post-task interviews confirm that these interpretations match subjects' experiences.

The inherent noisiness of the data make it difficult to quantitatively characterize these patterns. Nevertheless, a predictive model should attempt to reproduce them at least qualitatively.

**D. Thinking Times v. Term Relatedness**

The basic hypothesis underlying the design of the FTE task is that a subject's list of terms, together with the associated timing data, contains information about the subject's CKS. A subordinate hypothesis is that useful information can be extracted from the timing data itself, without reference to the meaning of the response terms. This is important for the eventual design of automated assessment/evaluation systems, and is significant for modeling that attempts to interpret FTE data.

This subsection relates FTE thinking times to one aspect of term meanings: the degree to which the meaning of a term in a subject's response list is related or unrelated to the meanings of the terms immediately preceding it in the list. According to the introspective testimony of experts who served as FTE subjects in preliminary studies, term entry events can be crudely classified into two types: those for which the term to be entered was immediately suggested by terms directly preceding it, and those for which the subject had to search his memory for some period of time to think of the term. According to these expert subjects, the immediately suggested terms were generally closely related in meaning to one of the prior few terms, while the terms thought of after a period of mental searching were usually related distantly or not at all to prior terms. This suggests the following hypothesis: let the term *jump* refer to a term which is "relatively unrelated" to any one of the previous few terms in an FTE response list; then, in FTE response data, longer thinking times should be statistically more likely to occur for jumps, while short thinking times should be more likely to occur for non-jumps.

To make this hypothesis testable, "relatively unrelated" and "previous few terms" must be defined. A neighborhood of three preceding terms was used ($n = 3$), rather than the one immediately preceding term, because interviews and perusal of term lists suggested that subjects often enter a term and then enter a sequence of multiple terms that come to mind approximately simultaneously. That is, a subject enters term A, and almost immediately thinks of terms B and C which are related to A but not necessarily related to each other; the subject then enters B and C. C is therefore a jump if one considers it relative to B only, but not if A is also part of the context. It is the testimony of some expert subjects that sometimes when they think of a term they perceive a "fork" in the mental path, with two possible "threads" of related terms that they could follow. In such a case they often try to follow one thread while it is productive, and then return to the fork and pick up the other thread. It seems reasonable that if the first thread is longer than about three terms, remembering and returning to the other thread is likely to require a pause for reflective thinking, thus sharing the characteristics of a jump and warranting classification as such.

Each term in each response list analyzed was compared to the three previous responses in the list by one domain expert (the first author), and subjectively declared to be a jump or a non-jump. Making explicit the criteria used to judge whether any pair of terms is related or not is difficult. Experts seem to possess an intuitive notion of whether terms are related, but have difficulty explicitly generalizing their criteria, probably because terms can have many possible kinds of relationships. In addition, experts tend to use contextual information in their judgments, inferring what the subject was thinking while entering a series of terms and deciding whether a term is a jump in that context. The following list specifies some situations in which two terms would be considered related (i.e., a non-jump):

- They were both within a sufficiently narrow topic area (e.g., collision-related terms, or graph-related terms, or angular momentum terms);
- They were analogous elements of a set or list (e.g., kinds of forces, or units of measure, or key principles of mechanics);
- One was a subclass or special case of the other (e.g., "force" and "spring [force]", or "motion" and "rotation");
- One was a situation or problem type in which the other figures significantly (e.g., "falling objects" and "gravity", or "collision" and "impulse");
- They were mathematically related (e.g., "work" and "impulse", or "velocity" and "position"); or
- One was a feature or element of the other (e.g., "slope" and "graph", or "force" and "free-body diagram").

This is not a complete list, but it illustrates the kinds of relationships considered.

Note that a very important question has been ignored so far: related *to whom*? The original hypothesis was that long thinking times correlate with terms unrelated to immediately preceding terms *according to the subject's own knowledge structure*. When an expert analyzing a subject's data identifies terms as jumps or non-jumps, however, the judgment of relatedness is made according to the expert's understanding of the domain, not the subject's. So, even if the hypothesis is completely correct and thinking times correlate perfectly with jumps, analysis by an expert would not show a perfect correlation unless the expert and subject were in complete agreement about what terms are and are not strongly related. We assume, however, that an expert with experience teaching the domain material can make judgments based on a structure that is reasonably close to what earnest students, or at least the more apt ones, possess. With that assumption, the operational hypothesis to test is that long thinking times will correlate noisily but significantly with jumps as perceived by an expert. Having subjects judge the relatedness of their own term pairs would be desirable, but requires research tools and techniques that are still being developed.

For each of the 16 subjects in the Physics I study, a domain expert reviewed the list of response terms for the FTE task and classified each term as a *jump* or *non-jump* as explained above. The set of thinking times for the subject's task performance was then divided into a subset containing thinking times for jumps and a subset containing thinking times for non-jumps. Figure 4 shows histograms of these two subsets for the example subject used above, superimposed on one set of axes. The natural logarithms of the thinking times have again been used rather than the times themselves. The histograms shown for this subject are typical, although some subjects' histograms are noisier with less clearly-defined peaks, and some have a greater fraction of terms in the "jump" category. For all but one of the 16 subjects, however, the mean and median of the jump distribution are clearly larger than the mean and median of the non-jump distribution. The lone exception is a data set containing atypically few points, resulting in atypically sparse, noisy histograms with similar means and medians. Most subjects have more non-jumps than jumps. The general pattern is clear: for any given subject, the thinking times associated with jumps are generally larger than the thinking times associated with non-jumps, though the two distributions overlap significantly.

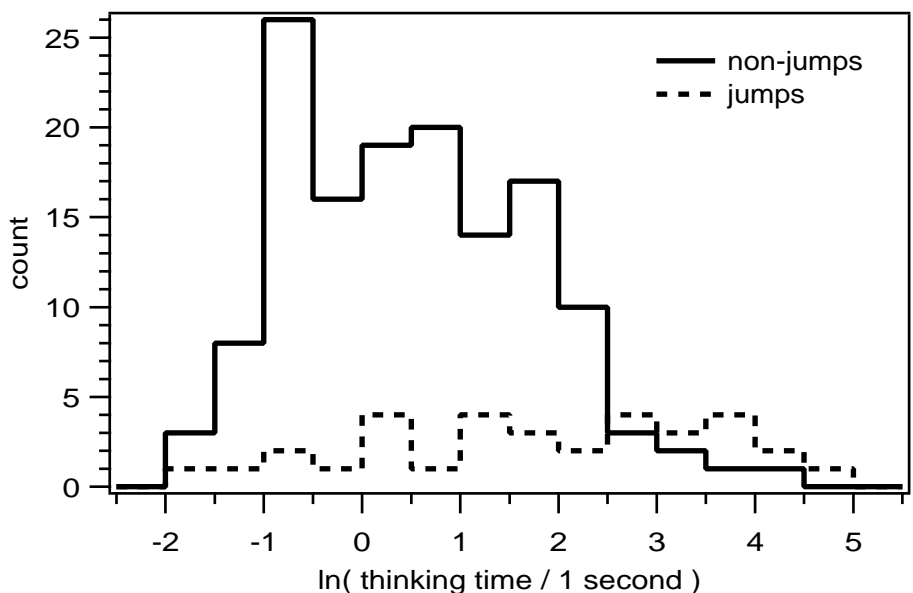

Figure 4: Comparison of histograms of logarithms of thinking times for jumps and for non-jumps, for subject 01 of the Physics I study.

This serves as confirmation that thinking time does correlate with term relatedness: on average, related terms (as inferred by an expert) come to mind more quickly than do unrelated terms. Had we used subjects' own judgment of term relatedness, the jump and non-jump histograms might have overlapped less and the correlation been stronger.

### IV. The Matrix Walk Model

The analysis of Section III indicated that Free Term Entry data have the following properties:
- The logarithms of the set of thinking times from an FTE data set approximately follow a normal distribution, with a pronounced leading spike;

- The rate of term entry gradually and unevenly decreases throughout task completion, as increasingly large thinking times appear, but tight clusters still appear; and
- A term's thinking time depends, at least in part, on how strongly the term is related to the previous term.

A model that purports to represent the cognitive processes underlying a subject's observed behavior on an FTE task should display these features.

In this section we present the *matrix walk model*, which is capable of displaying most of the listed features. The model is intended to be simple yet reasonable, relatively abstract, and quantitative. It is constructed in such a way that thinking times are entirely determined by term relationship strengths. It will be shown that the model can produce a distribution of thinking times that is approximately log-normal, and naturally exhibits a decreasing term entry rate throughout a simulated FTE task; the leading-edge spike found in real FTE data has not yet been recreated, however. Connecting the model to lower-level descriptions of cognition, such as neural network models (Hertz, Krogh, & Palmer, 1991; Hopfield, 1982) or production system models (Anderson, 1993), has not been attempted.

**A. Model Description**

The model represents a subject's conceptual knowledge store for a particular domain as an $N$ by $N$ *link matrix* **L** of real numbers, where $N$ is the number of *knowledge elements* in the structure. These knowledge elements are assumed to represent concepts or, equivalently, terms. The model does not ascribe a specific meaning like "energy" or "vector" to any of the knowledge elements; they are abstract. Each matrix element $L_{i,j}$ ($i \in \{1, 2, …, N\}$, $j \in \{1, 2, …, N\}$) represents the strength of the *link* from element $j$ to element $i$: the degree to which element $j$ "triggers" or "brings to mind" element $i$. **L** is not necessarily symmetric, and its diagonal elements are irrelevant.

The "dynamics" of the model is an algorithm called *the walk* which represents carrying out an FTE task, and which produces simulated FTE data analogous to the data collected from subjects. The algorithm is a set of rules for generating a list of the knowledge elements $\{a_k\}$ representing a response term list, given a specific matrix and a choice for the initial "prompt" term. A thinking time $\tau_k$ is determined for each element. The model does not address typing times.

The walk is defined by the following steps:

1. The first "active" element is arbitrarily chosen as element one: $a_0 = 1$. There is no associated thinking time. This element is ineligible for future selection.

2. Given a currently active element $a_k$, the next active element $a_{k+1}$ is chosen to be the one for which the link strength $L_{a_{k+1}, a_k}$ is maximal, excluding previously-active elements $a_0, a_1, …, a_k$. In other words, look at all the links leading from the current element, and follow the strongest which leads to an unvisited element.

3. The thinking time for the recall process of step 2 is defined to be $\tau_{k+1} = \tau(L_{a_{k+1}, a_k})$, where $\tau(s)$ is a recall function which will be discussed below. [In the notation used here, $\tau(s)$ represents a mathematical function, while $\tau_k$ represents one particular thinking time value.]

4. Update the "counter" variable $k \to k + 1$.

5. Go back to step 2 and repeat, unless a criterion for task termination (e.g., total number of terms) has been met, in which case the task is finished.

To complete the model, values for the matrix elements of **L** must be specified, and the recall function $\tau(s)$ must be defined.

**B. Choice of Recall Function**

Within the model, the thinking time associated with a particular FTE response element (term) is uniquely determined by the link strength to that element from the previously entered element. The recall function $\tau(s)$ determines the mapping between link strength value and thinking time. The following restrictions are imposed on the recall function:

- It should tend to small values (zero or some specified minimum possible thinking time) for link strengths approaching their maximum possible value;
- It should tend to infinity for link strengths approaching their minimum possible value; and
- It should be a relatively simple, well behaved mathematical function amenable to analysis.

The first two criteria reflect the notion that strong conceptual connections should produce quick responses, while weak connections should produce delayed responses. Since FTE thinking times in the study data were observed over more than two orders of magnitude, with vanishingly small frequencies of the largest times, having the recall function tend asymptotically to infinity is appropriate. The third criterion enforces parsimony and analytic convenience.

For link strengths between zero and one, two candidate recall functions are the *logarithmic recall function*

$$\tau(x) = -\alpha \ln(x) \qquad [1]$$

and the *power-law recall function*

$$\tau(x) = \alpha \left( x^{-\gamma} - \beta \right) \qquad [2],$$

where $\alpha$, $\beta$, and $\gamma$ are model parameters with ranges $\alpha > 0$, $0 < \beta \leq 1$, and $\gamma \geq 1$. For both functions, the parameter $\alpha$ sets the time scale. For the power law, $\beta$ determines the minimum possible thinking time, and $\gamma$ controls the relative abundance of long vs. short thinking times.

In the absence of *a priori* theoretical arguments for the form of the recall function (which might come from lower-level cognitive science principles), a particular recall function can only be chosen based on its simplicity and its success in helping the model reproduce observed data patterns.

**C. Amorphous Matrix**

A simple way to populate a link matrix and investigate the behavior of the model is to choose random numbers for all link strengths. This produces an "amorphous" or "structureless" matrix: one with no large-scale structure or pattern. Accordingly, link matrices were constructed by randomly drawing all link strengths from a distribution uniform between zero and one. The irrelevant diagonal elements were set to zero. The amorphous matrix is analytically convenient because it is possible to work backwards, deriving the recall function necessary to produce any desired distribution of (simulated) thinking times.

To simplify initial investigations, a log-normal distribution of thinking times was chosen as the goal since observed FTE data was crudely log-normal. (Predicting the leading spike seen in real FTE data with a more complicated, non-monotonic recall function was not attempted.) Determining a recall function that provides the appropriate distribution of thinking times is done by deriving an expression for the probability density function (PDF) of the strengths of the links "followed" by the model walk for a given distribution of link strengths, and determining the choice of recall function necessary to map that to the desired distribution of thinking times.

A simplification must be made to deal with the fact that the distribution changes as elements are chosen and the pool of "eligible" elements dwindles. It is assumed that whenever an element is chosen and made ineligible, a new element with new randomly-chosen link matrix elements replaces it. This is equivalent to assuming that the number of elements chosen during a complete FTE task is much smaller than the total number $N$ of available elements. Another way to look at it is that the distribution derived only describes the first thinking time of a task for each of an ensemble of subjects. Since the calculation is only intended to aid in the selection of a recall function and reasonable choices for the recall function's parameters, this approximation is not a major compromise.

The walk algorithm prescribes that an element is chosen as a response term by finding the $i$ for which $L_{i,j}$ has the largest value, given the previously chosen element $j$. For a random link matrix, this means choosing the largest of $N-1$ random numbers drawn independently from a distribution uniform between zero and one. The probability distribution for the result is therefore

$$p(s) = (N-1)\, s^{N-2} \qquad [3],$$

where $s$ is the value of the selected link strength (matrix element). This distribution has the distinctive feature that for large $N$ it weights strengths close to one extremely strongly and assigns vanishing probability to other values. The implications of this feature are important and will be discussed below.

Given a recall function $\tau(s)$, the distribution of thinking times $q(\tau)$ corresponding to the distribution of selected link strengths $p(s)$ is described by the relation

$$q(\tau)\, d\tau = |p(s)\, ds| \quad \Rightarrow \quad q(\tau) = p(s(\tau)) \left| \frac{d\,s(\tau)}{d\tau} \right| \qquad [4],$$

where $s(\tau)$ is the inverse of the recall function $\tau(s)$. If the desired distribution $q(\tau)$ is known, the recall function can be solved for by integrating Equation 4 to get

$$\int_0^\tau d\tau'\, q(\tau') = \int_s^1 ds'\, p(s') \qquad [5],$$

performing the integrals, and solving for $\tau$ in terms of $s$.

To produce thinking times with a log-normal distribution,

$$q(\tau) = \frac{1}{\sqrt{2\pi}\,\sigma\,\tau} \exp\left( \frac{-(\ln(\tau) - \mu)^2}{2\sigma^2} \right) \qquad [6]$$

where $\mu$ and $\sigma$ describe the mean and width of the peak, respectively, on a logarithmic plot. Inserting this and Equation 3 into Equation 5, integrating, and solving for $\tau$ results in

$$\tau(s) = \exp\left( \mu - \sigma\sqrt{2}\,\operatorname{erf}^{-1}\left( 2s^{N-1} - 1 \right) \right) \qquad [7]$$

where $\operatorname{erf}^{-1}(x)$ is the inverse of the error function $\operatorname{erf}(x)$. The inverse error function can be evaluated numerically but is analytically difficult, suggesting that one of the candidate recall functions presented in Equations 1 and 2 would be a better choice for the model if it could reasonably approximate Equation 7 with the right parameter choices.

The logarithmic and power-law recall functions of Equations 1 and 2 were compared to the derived function of Equation 7. To determine parameter values for $\alpha$, $\beta$, and $\gamma$, each was chi-squared fit to Equation 7 with $\mu$ and $\sigma$ chosen from a typical data set. Various values of $N \leq 10$ were used; $N \geq 200$ would be more realistic, but the range of values produced by the function for large $N$ is so extreme that numerical overflow problems result.

It was found that the power-law recall function can better approximate the log-normal-derived recall function than can the logarithmic recall function. Furthermore, the best-fit value of $\beta$ is effectively unity. Though these results hold for relatively small values of *N*, they seem likely to hold for larger *N* as well and suggest that the best choice for a model recall function would be the power-law. The power-law can apparently be simplified by fixing $\beta = 1$, which means that recall times can range from zero to infinity.

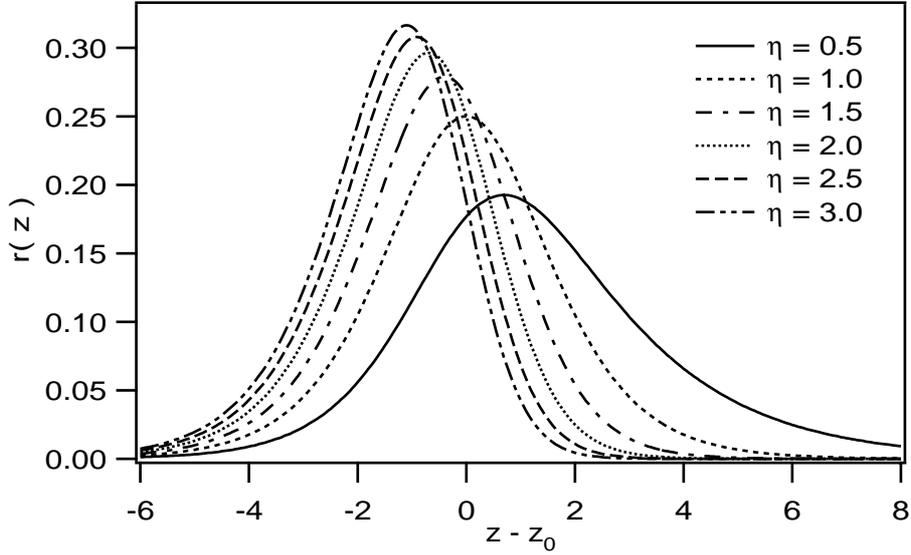

Figure 5: Probability distribution function r(z) for logarithms of thinking times generated by power-law recall function, for a range of values of the parameter $\eta$ and for $\beta = 1$.

Using Equation 4, the approximate distribution of early thinking times generated by the model (before element ineligibility becomes a significant effect) can be determined:

$$q(\tau) = \frac{N-1}{\alpha\,\gamma}\left(\frac{\tau}{\alpha}+\beta\right)^{-\left(\frac{N-1}{\gamma}+1\right)} \quad [8].$$

As discussed during analysis of the study data, it is more convenient to work with the logarithms of the thinking times. The distribution of thinking time logarithms for the model can be derived from Equation 8 with a simple change of variables, yielding

$$r(z) = \eta\, e^{z-z_0}\left(e^{z-z_0}+\beta\right)^{-(\eta+1)} \quad [9],$$

where $z \equiv \ln(\tau)$, $z_0 \equiv \ln(\alpha)$, and $\eta \equiv (N-1)/\gamma$. This distribution ought to resemble a Gaussian curve, and it does, as seen in Figure 5. With $\beta$ fixed at unity, the model only provides two parameters to control the shape of the distribution, since *N* and $\gamma$ always appear in the same combination (labeled $\eta$). $\alpha$ determines $z_0$, a horizontal axis offset (equivalent to setting the time scale for thinking times). $\eta$ controls the peak width and also impacts the location of the peak maximum, as demonstrated by Figure 5. With parameters to control both peak width and location, the distribution ought to be capable of modeling real FTE data except for the leading spike.

The model, with the power-law recall function, was used to generate synthetic FTE data. The distribution function of Equation 8 was used to estimate parameter values which would result in a distribution comparable to subject 01's distribution of thinking time logarithms (Figure 3). This subject was chosen because he demonstrated an unusually large number of response terms but appeared otherwise typical, resulting in less noisy data than that of most other subjects. Because the goal was to demonstrate that the model is capable of matching the general characteristics of FTE data, not to model the details of individual subjects, choosing one subject as an archetype introduces no compromises.

This provided initial guesses for $\eta$ and $\alpha$. Additional information is required to determine *N* and $\gamma$ from $\eta$. The choice of *N* affects how term entry rate changes as the task progresses: if *N* is much larger than the total number of terms *C* entered during the task, then the term entry rate does not change significantly; but if *N* is only slightly larger than *C*, term entry rate drops drastically, since only a few elements remain near the end. One can think of the parameter $\eta$ changing throughout the task, with

$$\eta_k \equiv \frac{N-1-k}{\gamma} = \eta_0 - \frac{k}{\gamma} \qquad [10]$$

being the value after *k* term entry events. ($\eta_0 = (N-1)/\gamma$.) As $\eta_k$ drops throughout the task, the distribution of resulting thinking times spreads and moves to higher values (see Figure 5), causing the rate of term entry to drop. Using this behavior to determine a reasonable estimate for *N* would require more precise data on FTE term entry rates than is currently available. Therefore, values of *N* in the range of 200-500 were explored numerically.

With initial parameter guesses in hand, the model was implemented on a computer and run to generate a synthetic data set. The simulation was terminated after 175 events had been generated, since the subject data being compared to consisted of 174 events. This was repeated many times for varying parameter values until a parameter set was found that produced data reasonably similar to the target subject's FTE data. More specifically, for each data set generated a quantile plot was constructed and fit with a log-normal cumulative distribution function; the parameters chosen consistently produced best-fit parameters close to those for a fit to the target subject's data. Since the intent was to demonstrate the model's general capability to produce reasonably realistic data, this was considered sufficient tuning of the parameters.

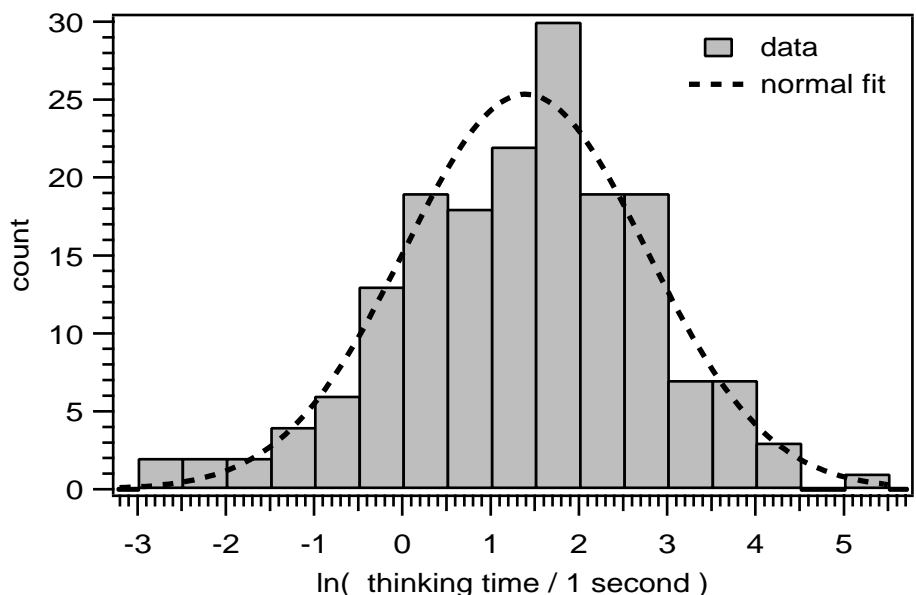

Figure 6: Histogram of logarithms of thinking times for model-generated data, using $N = 300$, $\gamma = 75$, $\alpha = 15$ sec, and $\beta = 1$. The probability density curve for the best-fit normal distribution is shown. (Compare to Figure 3.)

The final set of parameter values chosen was $N = 300$, $\gamma = 75$, and $\alpha = 15$ sec, with $\beta$ still fixed at 1. The resulting histogram for one instantiation (i.e., one randomly generated link matrix) is shown in Figure 6, along with a best-fit log-normal curve (actually a best-fit normal curve to the logarithms of the thinking times). For comparison, the equivalent plot for subject 01's FTE data can be seen in Figure 3. It is evident from these plots and from similar plots for other randomly generated link matrices that the model is capable of producing thinking time distributions that resemble real FTE thinking time distributions, though without the leading spike and slight skew seen in most subject's data. The model generally produces more outliers on both sides of the distribution than is seen in subject data. Low-end outliers could perhaps be eliminated by choosing a value of $\beta$ smaller than 1.

Figure 7 shows thinking times in order of occurrence for the same model-generated data (including randomly-generated typing times); for comparison, Figure 2 shows a similar plot for subject 01's data. The two plots are not unreasonably different in their gross characteristics, aside from one very long time in the model plot due to one of the aforementioned outliers. The model data does show a tendency towards decreasing term entry rate as the task progresses, and longer thinking times are significantly more likely to be found during the later part of the task.

The model appears relatively successful with an amorphous matrix. However, it is unsatisfying in one significant aspect, the size of the parameter $\gamma$ needed to produce a reasonable thinking time distribution. For plausible model output, it was found that $\eta = (N-1)/\gamma$ must be close to 4.0, and $N$ must be approximately 300, which requires $\gamma$ to be approximately 75, a huge value for an exponent. This can be understood in terms of distributions. The distribution of link strengths chosen according to the model's next-element selection algorithm is given by Equation 3. For large $N$, this distribution is astronomically heavily weighted in favor of values extremely close to one: the mean value of the distribution is $(N-1)/N$, which is 0.9967 for $N = 300$. This makes sense: for each walk step, the largest of a set of $N-1$ links is being selected. Because all strengths responsible for thinking times are therefore extremely close to one, with very little variation, a hypersensitive recall function is required to produce an acceptable range of thinking times. Thus $\gamma$, the exponent in the power law recall function of Equation 3, must be huge. This is unacceptable for a model that aspires to be cognitively interpretable, and may be interpreted as evidence that a matrix of randomly chosen numbers is not a good model of a subject's conceptual knowledge store and that other matrix-filling schemes should be investigated.

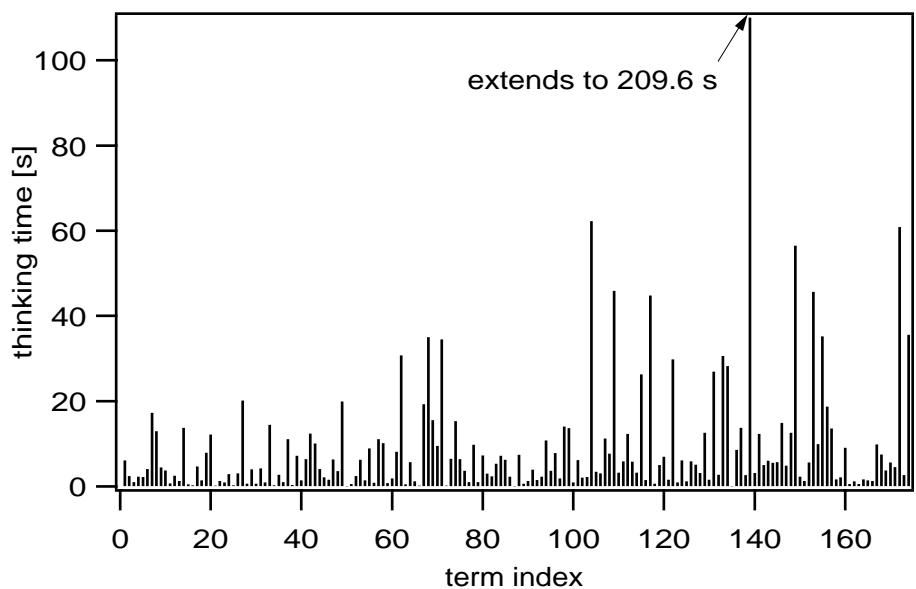

Figure 7: Thinking time vs. term index for model-generated data of Figure 6. (Compare to Figure 2.)

**D. Fractal Matrix**

Inspection of FTE responses from the studies shows that many terms entered by subjects are connected only tenuously if at all to the previous term, and analysis showed that these cases tend to correlate with larger thinking times. It therefore seems reasonable to insist on a model in which longer thinking times result from weak links, not from links almost equal to the strongest links. This suggests that the link matrix representing connections between knowledge elements should not be filled with numbers randomly drawn between zero and one, but according to a different and more structured scheme that forces the walk algorithm to select a broader range of link strengths. One possibility is to fill a randomly-chosen subset of the elements with numbers between zero and one, and set the rest to zero: a "sparse amorphous matrix". If the number of nonzero elements in each column is the same, defined to be $M$, then all of the preceding subsection's calculations for the amorphous matrix hold except for Equation 11, with $M$ replacing $N$. Effectively, this modification allows one to reduce the parameter $\gamma$ to a less problematic value without "running out of elements" too soon in the task by reducing $N$.

A different approach is to construct a matrix with a clustered structure. Qualitative results from physics education research suggest that a physics expert's knowledge structure is hierarchical, with strongly interlinked clusters of concepts, less strongly interlinked clusters of clusters ("metaclusters"), yet less strongly interlinked clusters of metaclusters, etc. (Gerace, Leonard, Dufresne, & Mestre, 1997). The resulting FTE "walk" through such a matrix should encounter all of the elements in a cluster, and then follow a weaker link to a new cluster within the same metacluster. Eventually, the metacluster would be exhausted and an even weaker link would be followed to a meta-metacluster. According to interviews, FTE subjects perceive themselves to enter a set of closely-related terms (e.g., on "circular motion" or "types of forces"), exhaust it, and then move on to another tightly-associated set.

One algorithm for generating a matrix with such a fractal structure is as follows. The algorithm requires that the number of knowledge elements $N$ be a power of 2. First, assign each matrix element a "zone" index as described in Figure 8; call the resulting matrix of zone indices $\mathbf{Z}$. Second, generate a "deterministic" matrix $\mathbf{D}$ of link strengths based on each element's zone:

$$D_{i,j} \equiv \mathrm{d}(Z_{i,j}) \qquad [11]$$

where

$$\mathrm{d}(z) \equiv \begin{cases} z^{-\rho} & (z \neq 0) \\ 0 & (z = 0) \end{cases} \qquad [12]$$

and $\rho$ is a model parameter that determines how weakly-linked metaclusters are relative to clusters. Third, generate a "random" matrix $\mathbf{R}$ of link strengths, in which diagonal elements are zero and off-diagonals are randomly selected from a distribution uniform between zero and one. Finally, combine the deterministic and random matrices into the final link matrix according to

$$\mathbf{L} \equiv (1-\lambda)\mathbf{D} + \lambda\mathbf{R} \qquad [13]$$

where $\lambda$ is a new model parameter that controls the degree of randomness in the link matrix. If $\lambda = 1$, the model reduces to the amorphous link matrix version studied in the previous subsection. When $\lambda \to 0$, the matrix limits to a regular, orderly fractal, although $\lambda$ cannot be strictly zero without causing the walk matrix to encounter ties in its decision-making step.

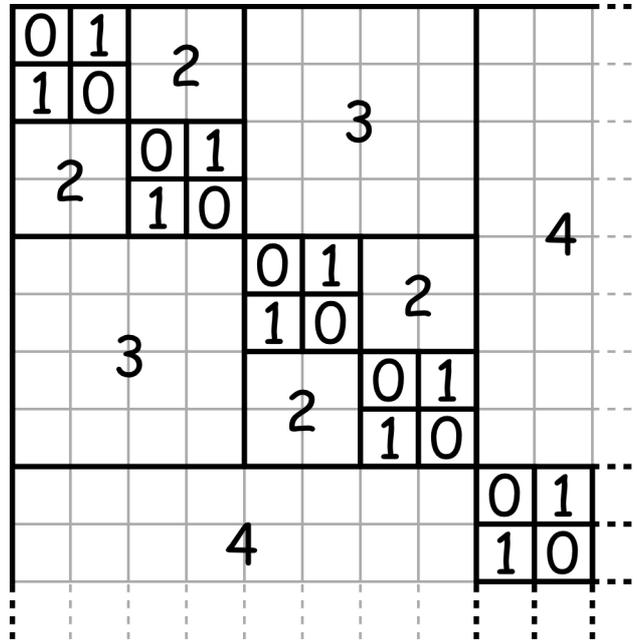

Figure 8: Zone number assigned to each matrix element for fractal algorithm.

This matrix is successful at distributing the link strengths selected by the walk across a wider range than is produced by the amorphous matrix. Figure 9 shows a plot of link strength selected vs. walk step index for one run of the model with $N = 256$, $\rho = 1$, and $\lambda = 0.01$. Clearly, the link strengths produced are distributed among several values between approximately 0.1 and 1, although they fall into distinct narrow bands with large gaps, and occur in a regular pattern as the walk progresses. This is a consequence of the regularity of the link matrix.

By increasing the randomness parameter $\lambda$, one can "smear out" the bands until they overlap, simultaneously decreasing the orderliness of the walk. Figure 10 shows a plot like that of Figure 9, but for one run of the model with $\lambda = 0.4$. The resulting link strengths are approximately 0.45 and greater. A histogram of the link strengths (not shown) is softly peaked around 0.65, and could be mapped onto a distribution of thinking times with a far less extreme recall function than was required for the amorphous matrix. This approach to generating a synthetic link matrix, with a combination of fractal-algorithm and random link strengths, thus appears to solve the large-$\gamma$ problem. The price paid is that the resulting distribution does not show a systematic decrease in term entry rate (decreasing link strengths) as is seen with the amorphous model and in the student data — a major flaw.

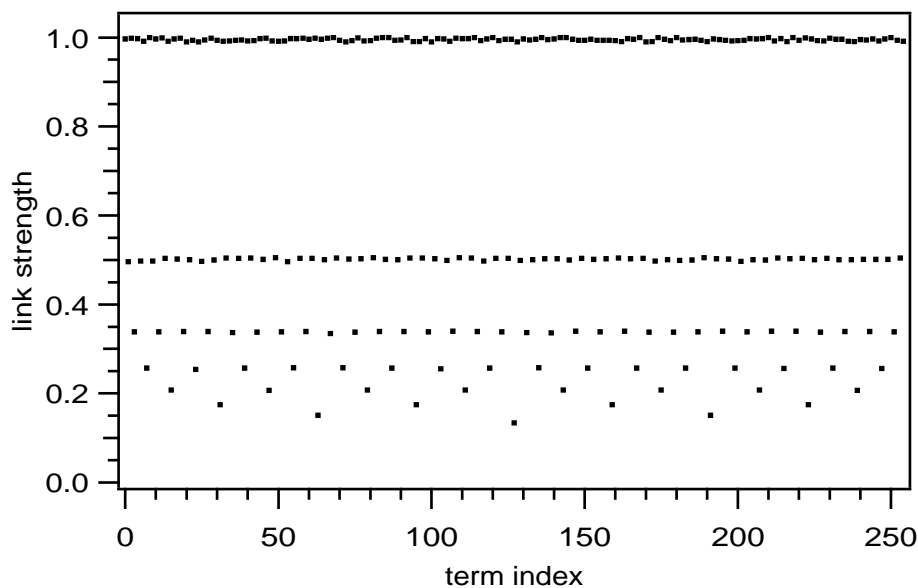

Figure 9: Link strengths vs. walk step index produced by the walk algorithm for a "fractal" matrix structure, with $N = 256$, $\rho = 1$, and $\lambda = 0.01$.

It is possible that a less regular and geometric algorithm for creating a hierarchical, clusters-of-clusters matrix with randomness in the sizes of clusters — a "random fractal" rather than a geometric fractal — would be superior in this regard. It would also better fit our preconceptions about the structure physics knowledge ought to assume in the mind, since we do not expect all topic and subtopic clusters to have exactly the same numbers of elements and links. It is also possible that the matrix should not describe an infinite regression of clusters and metaclusters, but merely two or three levels of clustering. Further investigations in this direction are presently underway.

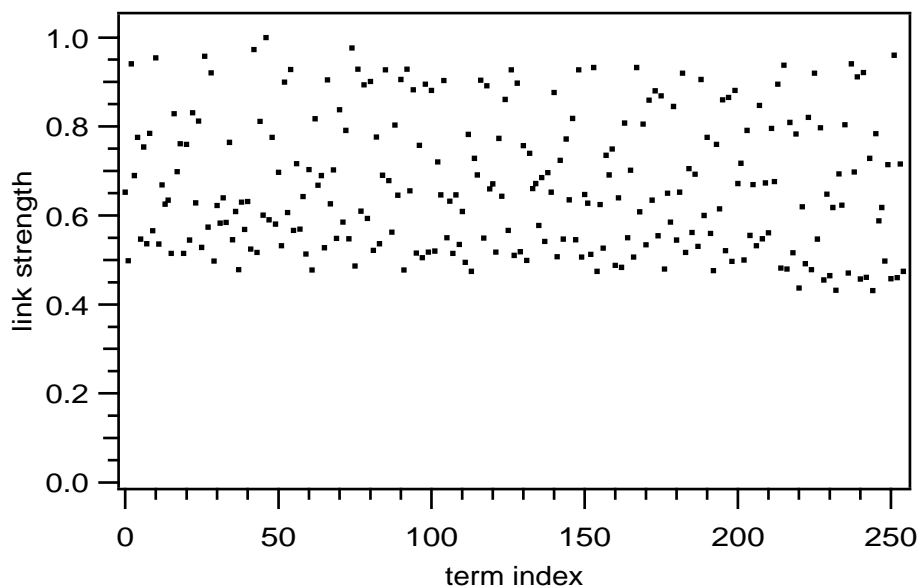

Figure 10: Same as Figure 9, but with $\lambda = 0.4$.

## V. Discussion

Section II of this paper defined and motivated the *Free Term Entry* task, a proposed instrument that is being investigated as a new approach to "cognitively diagnostic assessment". Section III described two distinctive phenomenological characteristics of observed FTE data: an approximately log-normal distribution of thinking time values which displays a leading spike and slight skew, and a general decrease in term entry rate as the task progresses. In addition, it presented evidence that the "thinking time" data captured by the FTE task depends at least partially on the strength of the relationship between term pairs in the subject's conceptual knowledge structure (CKS). These characteristics provide a target for modeling efforts.

Section IV proposed a simple, quantitative, dynamical model for how a subject accesses her CKS to generate observed FTE data. The fundamental components of the model are a "link matrix" describing the subject's CKS, a "recall function" indicating how the thinking time to remember a term depends on the term's association with the previous term, and a "walk" algorithm for selecting the sequence of knowledge elements (terms). The section investigated the model's ability to simulate FTE thinking times, using two different schemes for filling the link matrix. It was found that filling the matrix with random numbers could produce reasonable-looking output, although an unacceptably extreme recall function was required. Mixing the random matrix with a deterministic matrix having a geometric fractal structure allowed a more reasonable recall function, but lost the model's ability to demonstrate decreasing term entry rate.

The leading spike is a very interesting aspect of the observed data. Neither class of link matrix investigated here is capable of reproducing the spike without arbitrarily postulating a non-monotonic recall function. The spike might be a clue to a "realistic" link matrix structure, or an indication that the matrix walk model as proposed is fundamentally insufficient to explain FTE data. In general, a two-peaked distribution for a random variable suggests that two different mechanisms of production are involved, so perhaps two different recall processes are at work during FTE task completion.

Perhaps readers with a more extensive background in cognitive science can connect the model to established knowledge of cognitive mechanisms. A cognitive hypothesis for the spike-plus-peak distribution would be particularly useful, as would an *a priori* argument for a particular form of the recall function.

The FTE task is unlikely to be useful for practical assessment by itself, but could be valuable when combined with other, related tasks. One such task is the *Term Prompted Term Entry* (TPTE) task, in which students are presented with several different terms from a subject area, one at a time (Beatty, 2000; Beatty & Gerace, 2001). For each, they are asked to respond spontaneously with terms they consider "related" to the prompt term, until ten terms have been entered or ten seconds have passed without a term being entered. Where the FTE is intended to explore widely the concepts a subject associates with a subject and the general patterns of connectivity within them, the TPTE is designed to probe connections in more detail.

Consider an assessment method in which students are first given an FTE task to collect a vocabulary for a subject (and perhaps make some general deductions about the students' knowledge characteristics), and then given a battery of TPTE tasks with various terms they had responded with during the FTE and other TPTE tasks. In this way it might be possible and practical to construct a detailed, quantitative "map" of each students' conceptual knowledge structure for the subject.

As a research question, it would be interesting to gather many such maps for students and "expert" physicists of various levels, and look for general features and patterns which correlate with subject mastery. The connectivity patterns found could also suggest appropriate link-matrix construction algorithms for the Matrix Walk Model.

The kind of assessment described above could clearly be defeated by students that primed themselves by memorizing vocabulary lists beforehand. The approach would therefore not be useful for high-stakes "summative" assessment for assigning grades and determining placements and admissions, but rather for ongoing "formative" assessment to guide class-to-class teaching and learning.

Overall, the work described herein is not intended as a definitive presentation of conclusions, either empirical or theoretical. Rather, it is meant to suggest a novel approach to the challenge of assessment. It is also meant to serve as a kind of "existence proof" of the plausibility of bringing quantitative cognitive modeling into practical physics education research, and ultimately into educational practice.